\documentclass[a4paper, 12pt, twoside]{article}
\usepackage{a4wide}
\usepackage[centertags]{amsmath}
\usepackage{amsfonts}
\numberwithin{equation}{section}
\newcommand{\DS}[1]{\hbox to 0pt{/\hss}#1}
\begin{document}
\begin{titlepage}
\begin{center}
\hspace*{10cm} BONN-TH-99-10\\
\hspace{10cm} hep-th/9907082\\
\hspace{10cm} July 1999\\
\vspace*{2cm}

{\bf \LARGE Minimal Off-Shell Supergravity in Five Dimensions}\\
\vspace{1cm}

{\sc Max Zucker}\footnote{e-mail: zucker@th.physik.uni-bonn.de}
\vspace{.2cm}

{\it { Physikalisches Institut \\
Universit\"at Bonn\\
Nussallee 12\\
D-53115 Bonn, Germany}}\\
\end{center}
\vspace{1.2cm}

\begin{center}
{\large Abstract}
\end{center}
\vspace{.3cm}

We give an off-shell formulation of $N=2$ Poincar\'e supergravity in five dimensions.

\end{titlepage}
\newpage
\section{Introduction}   
The construction of off-shell supergravity theories is up to date a complicated task. Although useful and in a limited sense powerful techniques have been developed, setting up a locally supersymmetric lagrangian is still a very painful work. Our motivation for such a tedious project is twofold. 

First, we note that off-shell formulations are known for $D=4$, $N=1$ \cite{Stelle:1978ye, Ferrara:1978em} and \cite{Sohnius:1981tp}, $D=4$, $N=2$ \cite{Breitenlohner:1980np, deWit:1980ug, Fradkin:1979cw} and for $D=6$, $N=2$ \cite{Bergshoeff:1986mz}. Further results are known on conformal supergravity, e.g. $D=4$, $N=4$ \cite{Bergshoeff:1981is}. Clearly, there is a gap in five dimensions. With this work we would like to contribute to its closure. Partial results in this direction were obtained in \cite{ Howe:1981ev,Howe:1981nz} where the $N=2$ and $N=4$ theories in five dimensions were analyzed using superspace methods. In particular, the field content of the current multiplet was derived and constraints for the superspace torsion were suggested. A somehow related work is \cite{Breitenlohner:1979rq}. Although these papers explain how to construct the theory in question, explicit formulae are missing. We will follow in large parts their suggestions and derive the complete off-shell supergravity theory. A more complete analysis which should include the development of a multiplet calculus is left for the future. 

Secondly, Ho\v{r}ava and Witten conjectured that the low energy limit of the strongly coupled heterotic $E_8\times E_8$ string is described by eleven dimensional supergravity compactified on $S^1/\mathbb{Z}_2$ \cite{Horava:1996qa, Horava:1996ma}. Anomaly cancellation requires ten dimensional Super Yang-Mills multiplets with gauge group $E_8$ on each boundary. In their papers Ho\v{r}ava and Witten construct the corresponding lagrangian as a perturbation series in $\kappa_{11}^{2/3}$, where $\kappa_{11}$ is the eleven dimensional gravitational coupling constant. They find that at the order $\kappa_{11}^{4/3}$ terms proportional to $\delta(x^{11})^2$ appear and so they terminate their expansion. Consequently, their lagrangian is not supersymmetric to all orders in $\kappa_{11}$. Motivated by their work, Mirabelli and Peskin investigated five dimensional Super Yang-Mills theory on $S^1/\mathbb{Z}_2$ coupled to chiral multiplets which live on the four dimensional boundaries of this orbifold \cite{Mirabelli:1998aj}. They developed an economic method for constructing completely supersymmetric actions on such spaces, at least for the globally supersymmetric case. Their  method heavily relies on off-shell formulations, that is, in order to generalize their construction to local supersymmetry an off-shell formulation of this theory is needed. Of course, it would be most interesting to apply there method to 11 dimensional supergravity. Unfortunately there is no off-shell formulation of this theory available so that the method of Mirabelli and Peskin is not applicable. More promissing seems the construction of the corresponding action in five dimensions, i.e. five dimensional supergravity compactified on $S^1/\mathbb{Z}_2$ where various matter multiplets live on the four dimensional boundaries. This action may be viewed as a toy model or, more ambitious, as a part of the effective five dimensional action as obtained by Lukas et al. by reducing the 11 dimensional theory to five dimensions \cite{Lukas:1998tt}.

The paper is organized as follows: In section \ref{mult1} we construct the neccessary multiplets. Starting from Super Yang-Mills theory we derive the multiplet of currents in section \ref{cur2}, from which the linearized minimal supergravity multiplet is deduced, section \ref{lin2} . In section  \ref{full1} these results are supercovariantized, yielding the full minimal multiplet. A further ingredient to our construction is the nonlinear multiplet which is constructed in section \ref{sect25}. We then move on and construct a physically sensible lagrangian, the main result of this paper, which is given in section \ref{seclag}. Short conclusions and comments are collected in section \ref{conc}. In the Appendix a detailed explanation of our notations and conventions is given.  

\section{Multiplets\label{mult1}}
 Let us start by analyzing the necessary degrees of freedom. This is most easily done by considering the outcome of a dimensional reduction from five to four dimensions. The on-shell supergravity multiplet in five dimensions contains the vielbein $e_m^a$, the gravitino $\psi_m$ and the graviphoton $A_m$. Reducing this multiplet one finds a four dimensional $N=2$ supergravity multiplet and an $N=2$ Maxwell multiplet. The necessary off-shell field content of these multiplets is easily found (see e.g. \cite{Breitenlohner:1981ej, Breitenlohner:1980np}): The $D=4, N=2$ supergravity multiplet contains $40+40$ components and the corresponding Super Yang-Mills multiplet contains $8+8$ components. So we conclude that the minimal off-shell supergravity in five dimensions requires at least $48+48$ components.

 However, as in four dimensions it is possible to find a smaller multiplet on which the supersymmetry algebra closes: the so called minimal multiplet. This four dimensional multiplet has been constructed in \cite{Breitenlohner:1981ej} using superspace methods. In five dimensions the linearized version of this multiplet is the one which couples minimally to the multiplet of currents. The latter is easily constructed and using the knowledge of the precise form of the coupling term one can easily deduce the linearized transformation laws of the minimal multiplet.  
Knowing the linearized transformation laws, the construction of the full transformation laws of the minimal multiplet is then straightforward. The fields belonging to this multiplet are listed in table \ref{table1}.
\begin{table}
 \begin{center}
 \begin{tabular}{ccccccc}
 \hline\hline
 Type & Field & $SU(2)$ & dimension & \multicolumn{3}{c}{components}\\
&&&&bosonic&&fermionic\\

 \hline
 vielbein & $e_m^a$ & \bf 1 & 0 & $25-5-10$&& \\
 graviphoton &  $A_m$ & \bf 1 & 3/2 & $5-1$&& \\
 gravitino & $\psi_m$ & \bf 2 & 2 &&& $40-8$ \\
 isotriplet & $\vec{t}$ & \bf 3 & 5/2 & 3&& \\
 antisymmetric tensor & $v_{ab}$ & \bf 1 & 5/2 & 10&&\\
 $SU(2)$ gauge field & $\vec{V}_m$ & \bf 3 & 5/2 & $15-3$&&\\
 spinor & $\lambda$ & \bf 2 & 3 &&& 8\\
 scalar & $C$ & \bf 1 & 7/2 & 1&&\\
 \hline\hline
 &&&&40&+&40\\
 \end{tabular}
 \end{center}
 \caption{Field content of the minimal multiplet in $D=5$. In the last two columns the gauge degrees of freedom are subtracted.}\label{table1}
 \end{table}
It is possible to give an invariant action for this multiplet. However, as expected from the counting of degrees of freedom, this action turns out to be unphysical. An example being the existence of  an auxilliary scalar $C$ of (mass-) dimension $7/2$ for which the action reads
 \[
 S\sim\frac{1}{\kappa}\int d^5x\, eC.
 \] 
Here $\kappa$ is the gravitational coupling with dimension $-3/2$ and $e=\det e_m^a$ is the determinant of the vielbein. Varying this action w.r.t. $C$ one finds as equation of motion the meaningless equation $\det e_m^a=0$.

The solution of this problem is similar to the solution of the same problem in four dimensional conformal supergravity \cite{deWit:1981tn}. The idea is simple: Construct a multiplet which is supersymmetric if a constraint of the form
\begin{equation}
C+\partial_aV^a+\ldots=0\label{const1}
\end{equation}
is satisfied. Here $V_a$ belongs to this multiplet. Take this constraint as defining relation for $C$ and replace $C$ everywhere using this equation. Now the resulting invariant may be interpreted as supergravity action. A multiplet with a constraint of this form is the nonlinear multiplet. Furthermore, it contains $8+8$ degrees of freedom, so that together with the minimal multiplet $48+48$ field components are provided, verifying the counting of degrees of freedom in the beginning of this section.
 \subsection{The multiplet of currents}\label{cur2}
In this section we construct the multiplet of currents in five dimensions. This multiplet has been discussed previously in superspace \cite{Howe:1981nz}, see also \cite{Howe:1981ev}.

We start from on-shell abelian Super Yang-Mills theory in five dimensions. This theory contains the photon $A_a$ (with field strength $F_{ab}=2\partial_{[a}A_{b]}$), the gaugino $\lambda$ and a scalar $\phi$. The lagrangian \cite{Mirabelli:1998aj}
 \begin{equation}
 {\cal L} = -\frac{1}{4}F_{ab}F^{ab}+\frac{1}{2}\partial_a\phi\partial^a\phi+\frac{i}{2}\bar{\lambda}\gamma^a\partial_a\lambda \label{lag0}
\end{equation}
is invariant under the supersymmetry transformations
\begin{equation}
\begin{split}
\delta A_a = \frac{i}{\sqrt{2}}\bar{\varepsilon}\gamma_a\lambda,\qquad \delta \phi  =  \frac{i}{\sqrt{2}}\bar{\varepsilon}\lambda\\
\delta \lambda = \frac{1}{2\sqrt{2}}\gamma^{ab}\varepsilon F_{ab}-\frac{1}{\sqrt{2}}\gamma^a\varepsilon\partial_a\phi,
\end{split}\label{traf2}
\end{equation}
assuming that the field equations
\begin{equation}
\partial^a\partial_a\phi=\DS{\partial}\lambda=\partial_a F^{ab}=0 \label{feq}
\end{equation}
hold (For our conventions see the appendix).

The Noether currents of this theory are easily constructed by standard methods. They are 
\begin{itemize}
\item the (symmetrized) energy-momentum tensor 
\[
\Theta_{ab}=-F_{ac}F_b{}^c+\frac{1}{4}\eta_{ab}F_{cd}F^{cd}+\partial_a\phi\partial_b\phi-\frac{1}{2}\eta_{ab}(\partial\phi)^2+\frac{i}{4}\bar{\lambda}(\gamma_a\partial_b+\gamma_b\partial_a)\lambda,
\]
\item the current associated with the $SU(2)$ automorphism of the supersymmetry algebra which acts on the gaugino only,
\[
\vec{J}_a=-\frac{1}{4}\bar{\lambda}\vec{\tau}\gamma_a\lambda
\]
and
\item the supercurrent, following from the invariance of the lagrangian (\ref{lag0}) under the supersymmetry transformations (\ref{traf2}), 
\[
J^a=-\frac{1}{\sqrt{2}}\left(\gamma^b\gamma^a\lambda\partial_b\phi+\frac{1}{2}\gamma^{bc}\gamma^a\lambda F_{bc}\right).
\]
\end{itemize}
These currents are conserved, i.e.
\[
\partial_a J^a=\partial_a\Theta^{ab}=\partial_a\vec{J}^a=0
\]
if the field equations (\ref{feq}) are satisfied.

It is now straightforward to construct the complete multiplet of currents. From \cite{Howe:1981nz} we know that the lowest dimensional component of this multiplet is a scalar $C'$ with dimension three. We start by writing down the most general ansatz bilinear in fields for which the dimensions match and which is gauge invariant, i.e. $C'\sim \phi^2$. Then acting repeatedly with the transformations (\ref{traf2}) on this field, always using the field equations (\ref{feq}), one recovers the complete multiplet:   
\begin{eqnarray*}
\delta C' & = & \frac{i}{2}\bar{\varepsilon}\zeta\\
\delta \zeta & = & \frac{i}{2}\vec{\tau}\varepsilon\vec{X}+\frac{1}{2}\gamma^{ab}\varepsilon w_{ab}-\frac{i}{2}\vec{\tau}\gamma^a\varepsilon\vec{J}_a-\gamma^a\varepsilon\partial_a C'\\
\delta\vec{X} & = & \frac{1}{2}\bar{\varepsilon}\vec{\tau}\gamma^aJ_a-\bar{\varepsilon}\vec{\tau}\DS{\partial}\zeta\\
\delta w_{ab} & = & \frac{i}{4}\bar{\varepsilon}\gamma_{abc}J^c-i\bar{\varepsilon}\gamma_{[a}\partial_{b]}\zeta\\
\delta\vec{J}_a & = & -\frac{1}{2}\bar{\varepsilon}\vec{\tau}J_a\\
\delta J^a & = & \gamma_b\varepsilon\Theta^{ab}+\varepsilon J_{(1)}^a+i\gamma^b\vec{\tau}\varepsilon\partial_b\vec{J}^a-\frac{1}{2}\varepsilon^{abcde}\gamma_b\varepsilon\partial_cw_{de}\\
\delta J^a_{(1)} & = & -\frac{i}{2}\bar{\varepsilon}\DS{\partial}J^a\\
\delta \Theta^{ab} & = & -\frac{i}{4}\bar{\varepsilon}\gamma^{ca}\partial_cJ^b+(a\leftrightarrow b).
\end{eqnarray*}
Besides the currents which we have already discussed, the remaining fields of the current multiplet are the scalar $C'$, a spinor $\zeta$, an $SU(2)$ triplet and Lorentz scalar $\vec{X}$ and an antisymmetric two form $w_{ab}$. 
These fields are defined by
\[
C'=\frac{1}{2}\phi^2,\qquad \zeta=\sqrt{2}\lambda\phi,\qquad \vec{X}=\frac{1}{4}\bar{\lambda}\vec{\tau}\lambda,\qquad w_{ab}=\phi F_{ab}-\frac{i}{8}\bar{\lambda}\gamma_{ab}\lambda,
\]
further
\[
J^a_{(1)}=-\frac{1}{8}\varepsilon^{abcde}F_{bc}F_{de}-\partial_b(F^{ab}\phi).
\]
Note that this quantity is conserved, $\partial\cdot J_{(1)}=0$.
\subsection{The linearized minimal multiplet}\label{lin2}
The next step is to require invariance of the linearized gravity-matter coupling term
\[
\kappa\int d^5x \left(\frac{1}{8}h_{mn}\Theta^{mn}+\frac{i}{4}\bar{J}^m\psi_m-4C'C-2i\bar{\zeta}\lambda-\frac{1}{2}w_{ab}v^{ab}+\vec{X}\vec{t}+\frac{1}{2\sqrt{3}}A_mJ^m_{(1)}+\frac{1}{4}\vec{J}^m\vec{V}_m\right),
\]
under supersymmetry. $h_{mn}$ is the linearized metric and the other fields belonging to the minimal multiplet were listed  before in table \ref{table1}. One finds
\begin{equation}
\begin{split}
\delta h_{mn}& = -i\bar{\varepsilon}\gamma_{m}\psi_{n}+(m \leftrightarrow n)\\
\delta\psi_m & =  -\frac{1}{4}\gamma^{pn}\varepsilon\partial_p h_{mn}+\frac{1}{2}\gamma_{mab}\varepsilon v^{ab}+2i\vec{\tau}\gamma_m\varepsilon\vec{t}-\frac{i}{2}\vec{\tau}\varepsilon \vec{V}_m+\frac{1}{\sqrt{3}}\gamma^n\varepsilon\partial_n A_m\\
\delta A_m & =  -\frac{\sqrt{3}i}{2}\bar{\varepsilon}\psi_m\\
\delta\vec{V}_m & = \bar{\varepsilon}\vec{\tau}\gamma^n\partial_n\psi_m-4\bar{\varepsilon}\vec{\tau}\gamma_m\lambda\\
\delta v^{ab} & = \frac{i}{4}\varepsilon^{abcde}\bar{\varepsilon}\gamma_c\partial_d\psi_e+2i\bar{\varepsilon}\gamma^{ab}\lambda\\
\delta\lambda & = \frac{1}{4}\gamma_a\varepsilon\partial_bv^{ab}+\varepsilon C-\frac{i}{2}\vec{\tau}\gamma^a\varepsilon\partial_a\vec{t}\\
\delta C & = -\frac{i}{2}\bar{\varepsilon}\gamma^a\partial_a\lambda\\
\delta\vec{t} & = \bar{\varepsilon}\vec{\tau}\lambda.
\label{lintraf}
\end{split}
\end{equation}
The conservation of the currents $\Theta^{ab}, \vec{J}_a, J_a$ and $J_{(1)}^a$ induces various gauge symmetries on the fields they are coupled to. Apart from the standard symmetries for the linearized supergravity fields
\begin{equation}
\delta h_{mn}=\partial_m\xi_n+(m\leftrightarrow n), \qquad \delta \psi_m=\partial_m\eta,\qquad \delta A_m= \partial_m \Lambda\label{sym22}
\end{equation}
there is an additional $U(1)^3$ which follows from the conservation of $\vec{J}^a$. As a consequence, $\vec{V}_m$ transforms as gauge field,
\begin{equation}
\delta \vec{V}_m=\partial_m\vec{\xi}.\label{u13}
\end{equation}
This $U(1)^3$ will be knitted together with the global automorphism $SU(2)$ at order $\kappa$ to give a local $SU(2)$. Note that $\vec{V}_m$ is an auxiliary field so that the automorphism group of the supersymmetry algebra will be gauged by an auxilliary field. The invariance under the second transformation in (\ref{sym22}) allows us to add a term
\[
\Delta \psi_m=-\partial_m(\frac{1}{\sqrt{3}}\gamma^n\varepsilon A_n)
\]
to the transformation law of the gravitino in (\ref{lintraf}) so that the last term in this expression becomes the graviphoton field strength. Similarly, we add a transformation of the form (\ref{u13}) to the tranformation law of $\vec{V}_m$ in (\ref{lintraf}) to complete the first term to the linearized gravitino field strength.

The multiplet which we have found here contains the $40+40$ components mentioned in the beginning of this section.

\subsection{The full minimal multiplet}\label{full1}
Next, we let the supersymmetry parameter become a local quantity. Consequently we covariantize the transformation laws (\ref{lintraf}) w.r.t. local Lorentz transformations, local $SU(2)$ transformations and finally supercovariantize these formulas. Then we calculate the algebra on the fields. The closure of the algebra requires various additional terms of order $\kappa$ and higher. The transformation rules we find are
\begin{eqnarray*}
\delta e_m^a& = &-i\kappa\bar{\varepsilon}\gamma^a\psi_{m}\\
\delta\psi_m &= & \frac{1}{\kappa}{\cal D}_m\varepsilon+\frac{1}{2}\gamma_{mbc}\varepsilon v^{bc}+2i\vec{\tau}\gamma_m\varepsilon\vec{t}-\frac{1}{\sqrt{3}}\gamma^n\varepsilon\hat{F}_{mn}\\
\delta A_m & = & -\frac{\sqrt{3}i}{2}\bar{\varepsilon}\psi_m\\
\delta\vec{V}_m& = & 2\bar{\varepsilon}\vec{\tau}\gamma^n\hat{\cal R}_{nm}-4\bar{\varepsilon}\vec{\tau}\gamma_m\lambda -2\kappa\bar{\varepsilon}\vec{\tau}\gamma^{ab}\psi_m v_{ab} + \frac{\kappa}{\sqrt{3}}\bar{\varepsilon}\vec{\tau}\gamma^{ab}\psi_m \hat{F}_{ab}-12i\kappa\bar{\varepsilon}\psi_m\vec{t}\\
\delta v^{ab} & = & \frac{i}{4}\bar{\varepsilon}\gamma^{abcd}\hat{\cal R}_{cd}+2i\bar{\varepsilon}\gamma^{ab}\lambda\\
\delta\lambda & = & \frac{1}{4}\gamma_a\varepsilon\hat{\cal D}_bv^{ab}+\varepsilon C-\frac{i}{2}\vec{\tau}\gamma^a\varepsilon\hat{\cal{D}}_a\vec{t}-\frac{i\kappa}{2\sqrt{3}}\vec{\tau}\gamma^{ab}\varepsilon\vec{t}\hat{F}_{ab}\\ & + & i\kappa\gamma^{ab}\vec{\tau}\varepsilon v_{ab}\vec{t}+\frac{\kappa}{48}\varepsilon^{abcde}\gamma_a\varepsilon\hat{F}_{bc}\hat{F}_{de}\\
\delta C & = & -\frac{i}{2}\bar{\varepsilon}\gamma^a\hat{\cal{D}}_a\lambda-\frac{3i\kappa}{4} \bar{\varepsilon}\gamma^{ab}\lambda v_{ab}-11\kappa\bar{\varepsilon}\vec{\tau}\lambda\vec{t}-\frac{\kappa}{2}\bar{\varepsilon}\vec\tau\gamma^{ab}\hat{\cal R}_{ab}\vec{t}\\
\delta\vec{t} & = & \bar{\varepsilon}\vec{\tau}\lambda.
\end{eqnarray*}
${\cal D}_m$ is covariant w.r.t. local Lorentz transformations and local $SU(2)$ transformations,
\[
{\cal D}(\hat{\omega})_m\psi_n=(\partial_m+\frac{1}{4}\hat{\omega}_m{}^{ab}\gamma_{ab}-\frac{i\kappa}{2}\vec{\tau}\vec{V}_m)\psi_n,
\]
and the spin connection in the covariant derivative is always supercovariant, if not mentioned otherwise. Hatted symbols refer to supercovariantized quantities and 
\begin{equation}
\hat{\cal R}_{mn}= {\cal D}_{[m}\psi_{n]}+\frac{\kappa}{2}\gamma_{ab[m}\psi_{n]}v^{ab}+2i\kappa\gamma_{[m}\vec{\tau}\psi_{n]}\vec{t}-\frac{\kappa}{\sqrt{3}}\gamma^a\psi_{[n}\hat{F}_{m]a}\label{grfist1}
\end{equation} 
is the supercovariantized gravitino field strength. Explicit expressions for the spin connection and its supercovariantization may be found in the appendix. Let us give a useful formula: the transformation law for the supercovariant spin connection is
\[
\delta \hat\omega_{mab}=-i\kappa(2\bar{\varepsilon}\gamma_{[a}\hat{\cal R}_{b]m}-\bar\varepsilon\gamma_m\hat{\cal R}_{ab}) +2i\kappa^2(2i\bar{\varepsilon}\vec{\tau}\gamma_{ab}\psi_m\vec{t}+\frac{1}{\sqrt{3}}\bar{\varepsilon}\psi_m\hat{F}_{ab}+\frac{1}{2}\bar{\varepsilon}\gamma_{abcd}\psi_mv^{cd}).
\]

The commutator of two supersymmetry transformations is
\begin{equation}\label{algebra}
\begin{split}
[\delta_Q(\eta),\delta_Q(\varepsilon)] & =  \delta_Q(i\kappa\bar{\varepsilon}\gamma^m\eta\psi_m)+\delta_{g.c.}(i\bar{\eta}\gamma^m\varepsilon)+\delta_{U(1)}(-\frac{\sqrt{3}i}{2\kappa}\bar{\varepsilon}\eta+i\bar{\varepsilon}\gamma^m\eta A_m)\\ & +  \delta_{Lt}(i\bar{\eta}\gamma^n\varepsilon\hat{\omega}_{nab} - 4\kappa\bar{\eta}\vec{\tau}\gamma_{ab}\varepsilon\vec{t}+\frac{2i}{\sqrt{3}}\kappa\bar{\eta}\varepsilon\hat{F}_{ab}+i\kappa\bar{\eta}\gamma_{abcd}\varepsilon v^{cd})\\ & +  \delta_R(-i\kappa\bar{\eta}\gamma^m\varepsilon\vec{V}_m+12i\kappa\bar{\eta}\varepsilon\vec{t}+2\kappa\bar{\eta}\vec{\tau}\gamma_{ab}\varepsilon v^{ab}-\frac{\kappa}{\sqrt{3}}\bar{\eta}\vec{\tau}\gamma^{ab}\varepsilon\hat{F}_{ab}),
\end{split}
\end{equation}
where $\delta_Q, \delta_{g.c.}, \delta_{Lt}, \delta_R, \delta_{U(1)}$  represent supersymmetry, general coordinate, local Lorentz, local $SU(2)$ and local\footnote{The graviphoton is the gauge field for the $U(1)$ and the only field which transforms under this symmetry.} $U(1)$ transformations, respectively. The expressions in the brackets give the parameters of the transformations in question. 

We have calculated the algebra on $e_m^a, \psi_m, A_m, \vec{V}_m, v_{ab}, \vec{t}$ and $\lambda$. We have not verified the closure on $C$. But from the algebra on the other fields it follows, that no freedom to add further terms to the transformation law of any field is left. 

It is now relatively easy to write down an invariant:
\[
S=\int d^5x\, e{\cal L},
\]
with
\begin{equation}\label{lag1}
\begin{split}
{\cal L}  =  & -  \frac{4}{\kappa}C-2i\bar{\lambda}\gamma^m\psi_m+\kappa\bar{\psi}_m\vec{\tau}\gamma^{mn}\psi_n\vec{t}-\frac{1}{\sqrt{3}}F_{ab}v^{ab}\\ & -  \frac{\kappa}{6\sqrt{3}}\varepsilon^{abcde}A_aF_{bc}F_{de}+\frac{i\kappa}{8\sqrt{3}}\varepsilon^{abcde}\bar{\psi}_a\gamma_b\psi_cF_{de}-32\vec{t}\vec{t}.
\end{split}
\end{equation}
Of course, this can only be a first step towards a physical action since, as stated in the beginning of this section, the field equation for $C$ is the unphysical $\det e_m^a=0$. Further there is no kinetic term for the gravitino, etc. It is obvious that something like this must happen since the multiplet presented here contains only $40+40$ degrees of freedom whereas we know from the analysis in the beginning of this section that an adequate multiplet, i.e. one which allows for a physical lagrangian must contain (at least) $48+48$ components. This is very similar to the result in four dimensions \cite{Breitenlohner:1981ej}.
\subsection{The nonlinear multiplet\label{sect25}}
To resolve the problems stated at the end of the last section we have to introduce a further multiplet, the nonlinear multiplet. The name comes from the fact that the transformation rules are nonlinear in the fields which belong to this multiplet. Our nonlinear multiplet is the five dimensional Poincar\'e variant of the superconformal nonlinear multiplet in six and four dimensions described in \cite{Bergshoeff:1986mz, deWit:1981tn}.

The nonlinear multiplet contains $8+8$ components. They are listed in table \ref{table2}. Together with the $40+40$ components of the last section they will make up the supergravity multiplet.

\begin{table}
\begin{center}
\begin{tabular}{ccccccc}
\hline\hline
Type & Field & $SU(2)\times SU(2)'$ & dimension & \multicolumn{3}{c}{components}\\
&&&&bosonic&&fermionic\\
\hline
scalar & $\phi^i{}_\alpha$ & $({\mathbf 2},{\mathbf 2})$ & 0 & 3\\
spinor &  $\chi$ & $({\mathbf 2},{\mathbf 1})$ & 2 &&& 8 \\
scalar & $\varphi$ &$({\mathbf 1},{\mathbf 1})$ & 5/2 & 1 \\
vector & $V_a$ & $({\mathbf 1},{\mathbf 1})$ & 5/2 & $5-1$ \\
\hline\hline
&&&&8&+&8\\
\end{tabular}
\end{center}
\caption{Field content of the nonlinear multiplet. The subtracted degree of freedom of $V_a$ is due to the constraint (\ref{const2}).}\label{table2}
\end{table}
The scalar $\phi^i{}_\alpha$ transforms from the left under the local $SU(2)$ (acting on the index $i$) and we have introduced an additional global $SU(2)'$ under which $\phi^i{}_\alpha$ transforms from the right (acting on the index $\alpha$). $\phi^i{}_\alpha$ is an element of $SU(2)$ , i.e. denoting its hermitian conjugate by $\phi^\alpha{}_i$ we have \cite{deWit:1981tn}
\[
\phi^i{}_\alpha\phi^\alpha{}_j=\delta^i_j,\qquad \phi^\alpha{}_i\phi^i{}_\beta=\delta^\alpha_\beta,\qquad \phi^\alpha{}_i=\varepsilon_{ij}\varepsilon^{\alpha\beta}\phi^j{}_\beta.
\]
The $\phi^i{}_\alpha$ allows the conversion of local $SU(2)$ indices to global ones. It is needed to construct a gauge equivalent formulation where the local $SU(2)$ is broken explicitly as will be done in the next section.

The nonlinear multiplet is also the starting point for off-shell supergravity with gauged $SU(2)$, i.e. where $\vec{V}_m$ is replaced by a propagating field. This has been worked out for the four dimensional case in \cite{deWit:1980ib}. To construct this theory, the global $SU(2)'$ (acting on the index $\alpha$) is gauged by introducing a Super Yang-Mills multiplet \cite{ich1}. Furthermore, by identifying an $SO(2)$ subgroup of the gauged $SU(2)'$ with the $U(1)$ for which the graviphoton is the gauge field yields supergravity with a cosmological constant, i.e. {\it AdS} supergravity. The details are explained for the four dimensional case by de Wit et al. in \cite{deWit:1981tn}.

Counting components of the nonlinear multiplet one finds that a bosonic constraint is needed to get equal numbers of fermionic and bosonic components. The required constraint is of the form discussed in the beginning of section \ref{mult1}, eq.(\ref{const1}). 

We have constructed this multiplet by starting with an ansatz which is motivated from the literature \cite{deWit:1981tn, Bergshoeff:1986mz}:
\[
\delta \phi^i{}_\alpha \sim (\vec{\tau})^i{}_j\phi^j{}_\alpha\bar{\varepsilon}\vec{\tau}\chi,
\]
and a general ansatz for $\chi$ bilinear in $\chi$ and $\phi^i{}_\alpha$. Implementation of the algebra (\ref{algebra}) yields the remaining terms. Our result is
\begin{eqnarray*}
\delta \phi^i{}_\alpha & = & -i\kappa(\vec{\tau})^i{}_j\phi^j{}_\alpha\bar{\varepsilon}\vec{\tau}\chi\\
\delta \chi^i & = & \frac{1}{2}\gamma^a\varepsilon^i V_a -\frac{1}{2}\varepsilon^i \varphi -\frac{1}{2\kappa}\phi^i{}_{\alpha}\hat{\cal D}_a\phi^\alpha{}_j\gamma^a\varepsilon^j + 2i\kappa \chi^i\bar{\chi}\varepsilon+3i(\vec{\tau})^i{}_j\varepsilon^j\vec{t}\\ & + & \gamma^{ab}\varepsilon^i(-\frac{1}{4\sqrt{3}}\hat{F}_{ab}+\frac{1}{2}v_{ab})-\frac{i}{4}\kappa\gamma^{ab}\varepsilon^i\bar{\chi}\gamma_{ab}\chi\\
\delta\varphi & = & i\bar{\varepsilon}\gamma^a \hat{\cal D}_a\chi+i\bar{\varepsilon}_i\gamma^a\chi^j\phi^i{}_\alpha\hat{\cal D}_a\phi^\alpha{}_j-\frac{i}{2}\bar{\varepsilon}\gamma^{ab}\hat{\cal R}_{ab}\\ & - & 2i\bar{\varepsilon}\lambda+\frac{i\kappa}{2}\bar{\varepsilon}\gamma^{ab}\chi v_{ab}+\frac{i\kappa}{2\sqrt{3}}\bar{\varepsilon}\gamma^{ab}\chi\hat{F}_{ab}-i\kappa\bar{\varepsilon}\gamma^a\chi V_a-i\kappa\bar{\varepsilon}\chi\varphi\\
\delta V_a & = & i\bar{\varepsilon}\gamma_a{}^b\hat{\cal D}_b\chi+i\bar{\varepsilon}_i\gamma_a{}^b\chi^j\phi^i{}_\alpha\hat{\cal D}_b\phi^\alpha{}_j+i\bar{\varepsilon}_i\chi^j\phi^i{}_\alpha\hat{\cal D}_a\phi^\alpha{}_j\\ & - & \frac{i}{2}\bar{\varepsilon}\gamma_a{}^{bc}\hat{\cal R}_{bc} - i\bar{\varepsilon}\gamma^b\hat{\cal R}_{ab} - 2i\bar{\varepsilon}\gamma_a\lambda + i\kappa\bar{\varepsilon}\gamma^{b}\chi v_{ab}\\ & + & 2\kappa\bar{\varepsilon}\vec{\tau}\gamma_a\chi\vec{t}+\frac{i\kappa}{2\sqrt{3}}\bar{\varepsilon}\gamma_{abc}\chi\hat{F}^{bc}-i\kappa\bar{\varepsilon}\chi V_a-i\kappa\bar{\varepsilon}\gamma_{ab}\chi V^b-i\kappa\bar{\varepsilon}\gamma_a\chi\varphi.\\
\end{eqnarray*}
 Closure of the algebra on $\varphi$ and $V_a$ requires the supersymmetric constraint
\begin{equation}
\begin{split}
&C -\frac{1}{16\kappa}\hat{R}(\hat{\omega})_{ab}{}^{ab}-\frac{1}{4}\hat{\cal D}_aV^a-\frac{1}{8\kappa}\hat{\cal D}^m\phi^i{}_\alpha\hat{\cal D}_m\phi^\alpha{}_i-\frac{i\kappa}{2}\bar{\chi}\gamma^m\hat{\cal D}_m\chi + \frac{i\kappa}{4}\bar{\chi}\gamma^{ab}\hat{\cal R}_{ab}\\ & + 2i\kappa\bar{\chi}\lambda+5\kappa\vec{t}\vec{t} -\frac{\kappa}{24}\hat{F}_{ab}\hat{F}^{ab}+\frac{\kappa}{4} v_{ab}v^{ab} - \frac{i\kappa}{2}\bar{\chi}_i\gamma^m\chi^j\phi^i{}_\alpha\hat{\cal D}_m\phi^\alpha{}_j\\ & -\kappa^2\frac{i}{4}\bar{\chi}\gamma^{ab}\chi v_{ab} -\frac{i\kappa^2}{4\sqrt{3}}\bar{\chi}\gamma^{ab}\chi\hat{F}_{ab} +\frac{\kappa}{4} V_aV^a -\frac{\kappa}{4}\varphi^2=0,
\end{split}\label{const2}
\end{equation}
with the supercovariant curvature scalar
\[
\hat{R}(\hat{\omega})_{ab}{}^{ab}=R(\hat{\omega})_{ab}{}^{ab}-4i\kappa^2\bar{\psi}_m\gamma_n\hat{\cal R}^{mn}+4\kappa^3\bar{\psi}_m\vec{\tau}\gamma^{mn}\psi_n\vec{t}-\frac{2i\kappa^3}{\sqrt{3}}\bar{\psi}_m\psi_n\hat{F}^{mn}-i\kappa^3\bar{\psi}_m\gamma^{mnab}\psi_nv_{ab}
\]
and let us also give an explicit expression for the supercovariant derivative on $V_a$:
\begin{eqnarray*}
\hat{\cal D}_m V^a & = & \partial_m V^a +\hat{\omega}_m{}^{ab}V_b- i\kappa\bar{\psi}_m\gamma^{ab}\hat{\cal D}_b\chi-i\kappa\bar{\psi}_{mi}\gamma^{ab}\chi^j\phi^i{}_\alpha\hat{\cal D}_b\phi^\alpha{}_j-i\kappa\bar{\psi}_{mi}\chi^j\phi^i{}_\alpha\hat{\cal D}^a\phi^\alpha{}_j\\ & + & \frac{i\kappa}{2}\bar{\psi}_m\gamma^{abc}\hat{\cal R}_{bc} +i\kappa\bar{\psi}_m\gamma_b\hat{\cal R}^{ab} + 2i\kappa\bar{\psi}_m\gamma^a\lambda - i\kappa^2\bar{\psi}_m\gamma_{b}\chi v^{ab} - 2\kappa^2\bar{\psi}_m\vec{\tau}\gamma^a\chi\vec{t}\\ & - & \frac{i\kappa^2}{2\sqrt{3}}\bar{\psi}_m\gamma^{abc}\chi\hat{F}_{bc}+i\kappa^2\bar{\psi}_m\chi V^a+i\kappa^2\bar{\psi}_m\gamma^{ab}\chi V_b + i\kappa^2\bar{\psi}_m\gamma^a\chi\varphi.
\end{eqnarray*}
\section{Lagrangians}\label{seclag}
It is now easy to write down a physically sensible action. We take the constraint of the last section, eq. (\ref{const2}), as the defining relation for the scalar $C$ and replace this field by the constraint everywhere it appears. Doing this with the lagrangian (\ref{lag1}) one obtains after partial integration
\begin{equation}\label{lagg1}
\begin{split}
{\cal L} = - & \frac{1}{4\kappa^2}R(\hat{\omega})_{ab}{}^{ab}+\frac{i\kappa}{8\sqrt{3}}\bar{\psi}_a\gamma^{abcd}\psi_b(2\hat{F}_{cd}-F_{cd})-\frac{i\kappa}{4}\bar{\psi_a}\gamma^{abcd}\psi_bv_{cd}\\ + & \frac{3i\kappa}{2}\bar{\psi}_a\psi_bv^{ab}-\frac{1}{6}\hat{F}_{ab}\hat{F}^{ab}-\frac{1}{\sqrt{3}}\hat{F}_{ab}v^{ab}+v_{ab}v^{ab}-\frac{\kappa}{6\sqrt{3}}\varepsilon^{abcde}A_aF_{bc}F_{de}\\ - & \frac{i}{2}\bar{\psi}_a\gamma^{abc}\hat{\cal R}_{bc}+8i\bar{\chi}\lambda+3\kappa\bar{\psi}_m\vec{\tau}\gamma^{mn}\psi_n\vec{t}+\frac{i}{2}\bar{\psi}_{mi}\gamma^{mna}\psi_n^j\phi^i{}_\alpha\hat{\cal D}_a\phi^\alpha{}_j\\ - & 6\kappa\bar{\chi}\vec{\tau}\gamma^m\psi_m\vec{t}+i\kappa\bar{\chi}\gamma^m\gamma^{ab}\psi_mv_{ab}-\frac{i\kappa}{2\sqrt{3}}\bar{\chi}\gamma^m\gamma^{ab}\psi_m\hat{F}_{ab}\\ - & i\kappa\bar{\psi}_a\gamma^a\gamma^b\chi V_b -i\kappa\bar{\psi}_m\gamma^m\chi\varphi+2i\bar{\chi}\gamma^{ab}\hat{\cal R}_{ab}-2i\bar{\chi}\gamma^m\hat{\cal D}_m\chi\\ - & 12\vec{t}^2 +V_aV^a-\varphi^2-i\kappa\bar{\chi}\gamma^{ab}\chi(v_{ab}+\frac{1}{\sqrt{3}}\hat{F}_{ab})\\ + & i\bar{\psi}_{ai}\gamma^{ab}\chi^j\phi^i{}_\alpha\hat{\cal D}_b\phi^\alpha{}_j+i\bar{\psi}_{ai}\chi^j\phi^i{}_\alpha\hat{\cal D}^a\phi^\alpha{}_j-\frac{1}{2\kappa^2}\hat{\cal D}_m\phi^i{}_\alpha\hat{\cal D}^m\phi^\alpha{}_i\\ - & 2i\kappa\bar{\chi}_i\gamma^m\chi^j\phi^i{}_\alpha\hat{\cal D}_m\phi^\alpha{}_j-\kappa^2\bar{\psi}_p\gamma^p\psi_m\bar{\chi}\gamma^{mn}\psi_n+\frac{\kappa^2}{2}\bar{\psi}_n\gamma^p\psi_m\bar{\chi}\gamma^{mn}\psi_p\\ - & \frac{\kappa^2}{4}\bar{\psi}_a\gamma^{abcd}\psi_b\bar{\chi}\gamma_{cd}\chi-\frac{\kappa^2}{2}\bar{\psi}_n\psi_m\bar{\chi}\gamma^{mn}\chi+2\kappa^2\bar{\psi}_n\gamma^{nm}\chi\bar{\chi}\psi_m.
\end{split}
\end{equation}
Note that the kinetic terms for the gravitino and the auxiliary spinor $\chi$ are non-diagonal. This may be remedied by defining
\begin{equation}
\lambda'=\lambda+\frac{1}{4}\gamma^{ab}\hat{\cal R}_{ab}.\label{lambda1}
\end{equation}
Furthermore, after replacing $v_{ab}$ by
\begin{equation}
v_{ab}'=v_{ab}-\frac{1}{2\sqrt{3}}\hat{F}_{ab},\label{vab1}
\end{equation}
the kinetic term for the graviphoton takes on its canonical form. Finally, the local $SU(2)$ symmetry is fixed by setting
\begin{equation}
\phi^i{}_\alpha=\delta^i_\alpha \label{gafix}
\end{equation}
and we arrive at a gauge equivalent formulation where the three degrees of freedom of the $\phi^i{}_\alpha$ are absorbed by the former $SU(2)$ gauge field $\vec{V}_m$. To maintain the gauge (\ref{gafix}) a field dependent $SU(2)$ transformation  with parameter 
\[
\vec{\xi}=2\kappa\bar{\varepsilon}\vec{\tau}\chi
\]
has to be added to the transfomation laws. This, however, introduces a term $\sim \partial\varepsilon$ into the supersymmetry transformation law of $\vec V_m$. We can get rid of this, by redefining
\begin{equation}
\vec{V}_m'=\vec{V}_m-2\kappa\bar{\psi}_m\vec{\tau}\chi.\label{vec1}
\end{equation}

To get some understanding of the lagrangian (\ref{lagg1}) we first decovariantize this expression, then we plug in the definitions (\ref{lambda1} -- \ref{vec1}) 
and finally split the lagrangian,
\begin{equation}
{\cal L} = {\cal L}_1+{\cal L}_2+{\cal L}_3.\label{lag22} 
\end{equation}
The on-shell part is given by \cite{Cremmer:1980gs, Chamseddine:1980sp}
\begin{equation}
\begin{split}
{\cal L}_1   =  - & \frac{1}{4\kappa^2}R(\tilde{\omega})_{ab}{}^{ab} -\frac{i}{2}\bar{\psi}_p\gamma^{pmn}{\cal D}(\frac{\hat{\omega}+\tilde{\omega}}{2})_m\psi_n - \frac{1}{4}F_{ab}F^{ab}\\ - & \frac{\kappa}{6\sqrt{3}}\varepsilon^{abcde}A_aF_{bc}F_{de}-\frac{\sqrt{3}i\kappa}{16}(F^{ab}+\hat{F}^{ab})\bar{\psi}_c\gamma^{[c}\gamma_{ab}\gamma^{d]}\psi_d
\end{split}
\end{equation}
and we use the same conventions as \cite{Dolan:1984fm}. On-shell $\tilde{\omega}_{mab}$ is the solution to its own field equation, \cite{Cremmer:1980gs}. An explicit expression is given in the appendix, eq. (\ref{spco1}). The covariant derivative is now only covariant w.r.t. local Lorentz transformations.

The off-shell lagrangians are:
\begin{eqnarray*}
{\cal L}_2 & = & v_{ab}'v^{{ab}'}+8i\bar{\chi}\lambda'-12\vec{t}\vec{t}+V_aV^a - \varphi^2-\frac{1}{4}\vec{V}'_a\vec{V}^{'a}+2i\kappa\bar{\chi}\gamma^m\gamma^{ab}\psi_mv_{ab}'+2i\kappa\bar{\chi}\psi_mV^m\\ & - & 2i\bar{\chi}\gamma^m{\cal D}_m\chi +\kappa\bar{\chi}\vec{\tau}\gamma^{mn}\psi_m\vec{V}'_n-i\kappa\bar{\chi}\gamma^{ab}\chi(v_{ab}'+\frac{\sqrt{3}}{2}\hat{F}_{ab})-12\kappa\bar{\chi}\vec{\tau}\gamma^m\psi_m\vec{t}
\end{eqnarray*}
and
\[
{\cal L}_3 = - \frac{\kappa^2}{4}\bar{\psi}_m\gamma^{mn}\gamma^{ab}\psi_n\bar{\chi}\gamma_{ab}\chi + 2\kappa^2\bar{\psi}_m\gamma^{mn}\chi\bar{\chi}\psi_n.
\]
The equations of motion for the auxilliary fields are
\[
v_{ab}'=\chi=\lambda'=\vec{t}=V_a=\varphi=\vec{V}'_m=0
\]
and on-shell only ${\cal L}_1$ remains.

For completeness we give the transformation laws under which the lagrangian (\ref{lag22}) is invariant. Explicit expressions for the supercovariant objects appearing here are listed in the appendix, eqns. (\ref{fs}, \ref{covder}). 
\begin{subequations}
\begin{eqnarray}
\delta e_m^a & = & - i\kappa\bar{\varepsilon}\gamma^a\psi_m\\
\delta A_m & = & - \frac{i\sqrt{3}}{2}\bar{\varepsilon}\psi_m\\
\delta \psi_m &  = & \frac{1}{\kappa}{\cal D}(\hat{\omega})_m\varepsilon +\frac{1}{4\sqrt{3}}\gamma_{mbc}\varepsilon\hat{F}^{bc}-\frac{1}{\sqrt{3}}\gamma^n\varepsilon\hat{F}_{mn}\nonumber\\ & - & \frac{i}{2}\vec{\tau}\varepsilon\vec{V}_m'+2i\gamma_m\vec{\tau}\varepsilon\vec{t}+\frac{1}{2}\gamma_{mbc}\varepsilon v^{'bc}+i\kappa\vec{\tau}\psi_m\bar{\varepsilon}\vec{\tau}\chi-i\kappa\vec{\tau}\varepsilon\bar{\chi}\vec{\tau}\psi_m\\
\delta \chi & = & - \frac{1}{2}\varepsilon\varphi+3i\vec{\tau}\varepsilon\vec{t}+\frac{1}{2}\gamma^a\varepsilon V_a -\frac{i}{4}\gamma^m\vec{\tau}\varepsilon\vec{V}_m'+\frac{1}{2}\gamma^{ab}\varepsilon v_{ab}' + i\kappa\chi\bar{\chi}\varepsilon\\
\delta \vec{t} & = & \bar{\varepsilon}\vec{\tau}\lambda'-\frac{1}{4}\bar{\varepsilon}\vec{\tau}\gamma^{ab}\hat{\cal R}_{ab}+2\kappa\vec{t}\times(\bar{\varepsilon}\vec{\tau}\chi)\\
\delta\varphi & = & i\bar{\varepsilon}\gamma^m\hat{\cal D}_m\chi-2i\bar{\varepsilon}\lambda' + \frac{i\kappa}{2}\bar{\varepsilon}\gamma^{ab}\chi v_{ab}'+\frac{i\kappa\sqrt{3}}{4}\bar{\varepsilon}\gamma^{ab}\chi\hat{F}_{ab}-i\kappa\bar{\varepsilon}\chi\varphi-i\kappa\bar{\varepsilon}\gamma^a\chi V_a\\
\delta V_a & = & i\bar{\varepsilon}\gamma_a{}^b\hat{\cal D}_b\chi-\frac{\kappa}{2}\bar{\varepsilon}\vec{\tau}\chi\vec{V}_a'-2i\bar{\varepsilon}\gamma_a\lambda' + i\kappa\bar{\varepsilon}\gamma^b\chi v_{ab}' + \frac{i\kappa}{2\sqrt{3}}\bar{\varepsilon}\gamma^b\chi\hat{F}_{ab}\nonumber\\ & + & 2\kappa\bar{\varepsilon}\vec{\tau}\gamma_a\chi\vec{t}+\frac{i\kappa}{2\sqrt{3}}\bar{\varepsilon}\gamma_{abc}\chi\hat{F}^{bc}-i\kappa\bar{\varepsilon}\gamma_a\gamma^b\chi V_b-i\kappa\bar{\varepsilon}\gamma_a\chi\varphi\\
\delta v^{'ab} & = & -\frac{i}{4}\bar{\varepsilon}\gamma^{[a}\gamma^{b]cd}\hat{\cal R}_{cd}+\frac{3i}{2}\bar{\varepsilon}\gamma^{[b}\gamma_c\hat{\cal R}^{a]c}+2i\bar{\varepsilon}\gamma^{ab}\lambda'\\
\delta \vec{V}'_a & = & -4\bar{\varepsilon}\vec{\tau}\gamma_a\lambda'+\bar{\varepsilon}\vec{\tau}\gamma_a{}^{bc}\hat{\cal R}_{bc}+2\bar{\varepsilon}\vec{\tau}\hat{\cal D}_a\chi + i\kappa\bar{\chi}\varepsilon\vec{V}_a' - \kappa\bar{\chi}\vec{\tau}\gamma_{abc}\varepsilon v^{'bc}\nonumber\\ & - & \frac{\kappa}{2\sqrt{3}}\bar{\chi}\vec{\tau}\gamma_{abc}\varepsilon\hat{F}^{bc}-4i\kappa\bar{\chi}\vec{\tau}\tilde{\tau}\gamma_a\varepsilon\tilde{t}+\frac{2\kappa}{\sqrt{3}}\bar{\chi}\vec{\tau}\gamma^b\varepsilon\hat{F}_{ab}-\kappa\bar{\varepsilon}\vec{\tau}\chi\times\vec{V}_a'\\
\delta \lambda' & = & \varepsilon\left(\frac{1}{4}\hat{\cal D}_aV^a+\frac{\kappa}{16}\vec{V}_a'\vec{V}^{'a}+\frac{i\kappa}{2}\bar{\chi}\gamma^m\hat{\cal D}_m\chi+\frac{i\kappa}{4}\bar{\chi}\gamma^{ab}\hat{\cal{R}}_{ab}- \frac{\kappa}{4}V_aV^a+15\kappa\vec{t}\vec{t}\right.\nonumber\\ & + & \left.\frac{i\kappa^2}{4}\bar{\chi}\gamma^{ab}\chi(v_{ab}'+\frac{\sqrt{3}}{2}\hat{F}_{ab}) +\frac{\kappa}{4}\varphi^2+ \frac{\kappa}{2}v_{ab}'v^{'ab}+\frac{\kappa}{2\sqrt{3}}v^{'ab}\hat{F}_{ab} - 2i\kappa\bar{\chi}\lambda'\right) \nonumber\\ & - & \frac{1}{2}\gamma_a\varepsilon(\hat{\cal D}_bv^{ab'}-\frac{\kappa}{2}\varepsilon^{abcde}v_{bc}'v_{de}'-\frac{\kappa}{\sqrt{3}}\varepsilon^{abcde}\hat{F}_{bc}v_{de}') + \gamma^{ab}\vec{\tau}\varepsilon\left(-\frac{i\kappa}{16}(\vec{V}_a'\times\vec{V}_b')\right.\nonumber\\ & - & 2i\kappa v_{ab}'\vec{t} + \left.\frac{i\sqrt{3}\kappa}{2}\hat{F}_{ab}\vec{t} -\frac{i}{8}\hat{\cal D}_a\vec{V}_b'-\frac{i\kappa}{4}\bar{\chi}\vec{\tau}\hat{\cal R}_{ab}\right) + \frac{3i}{2}\gamma^a\vec{\tau}\varepsilon\hat{\cal D}_a\vec{t}+\frac{3i\kappa}{2}\gamma^a\vec{\tau}\varepsilon (\vec{V}_a'\times t)\nonumber\\ & + & \frac{1}{4}\gamma^{abc}\varepsilon\hat{\cal D}_a v_{bc}' +  i\kappa\vec{\tau}\lambda'\bar{\varepsilon}\vec{\tau}\chi- \frac{\kappa}{\sqrt{3}}\gamma^{ab}\varepsilon\hat{F}_{ca}v^{'c}{}_b.
\end{eqnarray}\label{ftl}
\end{subequations}
The covariant derivatives are now only covariant w.r.t. local Lorentz transformations and the supercovariant gravitino field strength has become
\begin{align*}
\hat{\cal R}_{mn} = & {\cal D}_{[m}\psi_{n]}-\frac{i\kappa}{2}\vec{\tau}\psi_{[n}\vec{V}_{m]}'+\frac{\kappa}{2}\gamma_{[m}{}^{ab}\psi_{n]}(v_{ab}'+\frac{1}{2\sqrt{3}}\hat{F}_{ab})\\ + & 2i\kappa\gamma_{[m}\vec{\tau}\psi_{n]}\vec{t}-\frac{\kappa}{\sqrt{3}}\gamma^p\psi_{[n}\hat{F}_{m]p}-i\kappa^2\vec{\tau}\psi_{[n}\bar{\psi}_{m]}\vec{\tau}\chi.
\end{align*}
\section{Conclusions}\label{conc}
In this paper we have constructed minimal off-shell Poincar\'e supergravity in five dimensions. We started from a globally supersymmetric on-shell theory from which we constructed the associated multiplet of currents. By dualization we were able to deduce the linearized transformation laws of the minimal multiplet. We then constructed the full transformation laws which are very similar to the result of Breitenlohner and Sohnius \cite{Breitenlohner:1981ej}. As already found in this work such multiplets are not suited for the construction of physical actions. Extending their work we resolved this problem using the nonlinear multiplet, which is known from $N=2$ conformal supergravity in four dimensions \cite{deWit:1981tn}. Our final lagrangian is (\ref{lag22}) and the transformation laws under which it is invariant are given in eqns. (\ref{ftl}).

\subparagraph{Acknowledgments:} It is a pleasure to thank P. Breitenlohner and A. van Proeyen for helpful comments. Furthermore I would like to thank J. Conrad, H.-P. Nilles and especially S. F\"orste.

This work was partially supported by the European Commission programs ERBFMRX-CT96-0045 and CT96-0090. 
\begin{appendix}
\section{Notations and conventions}\label{app}
Our metric is $\eta_{ab}=diag(+----)$ and  $a,b=0,\ldots 4$ are Lorentz indices whereas $m,n=0,\ldots 4$ are world indices . The Dirac matrices satisfy the Clifford algebra
\[
\{\gamma^a,\gamma^b\}=2\eta^{ab}.
\]
Antisymmetrization is with strength one, i.e.
\[
\gamma^{(n)}=\gamma^{a_1\ldots a_n}=\gamma^{[a_1}\gamma^{a_2}\ldots\gamma^{a_n]}=\frac{1}{n!}\sum_{perm's}(-1)^p\gamma^{a_1}\ldots\gamma^{a_n}
\]
and the sum is over all permutations. The gamma matrices are chosen to satisfy
\[
\gamma^{a_1\ldots a_5}=\varepsilon^{a_1\ldots a_5},
\]
where the Levy-Civita symbol is totally antisymmetric and (with flat indices)
\[
\varepsilon^{01234}=\varepsilon_{01234}=1.
\]
Now we can give the duality relation for the gamma matrices:
\[
\gamma^{a_1\ldots a_p}=\frac{\alpha_p}{(5-p)!}\varepsilon^{a_1\ldots a_pb_1\ldots b_{5-p}}\gamma_{b_1\ldots b_{5-p}}, \qquad \alpha_p= \left\{ \begin{array}{ll}
  -1&\mbox{ for }p=2,3 \\
  +1&\mbox{ for }p=0,1,4,5
\end{array}\right..
\]
Spinors carry an $SU(2)$ index $i=1,2$. The Dirac conjugate is defined by
\[
\bar\psi_i\equiv(\psi^i)^\dagger\gamma^0.
\]
We can impose a symplectic Majorana condition,
\[
\bar\psi_i=\varepsilon_{ij}(\psi^j)^TC,
\]
where the charge conjugation matrix satisfies 
\[
C=-C^T=-C^{-1},\qquad \gamma_a^T=C\gamma_aC^{-1}.
\]
If possible we suppress the $SU(2)$ indices. In this case the summation of these indices is from southwest to northeast, e.g.
\begin{equation}
\bar{\psi}\gamma^a\lambda\equiv\bar{\psi}_i\gamma^a\lambda^i,\qquad \bar{\psi}\vec{\tau}\gamma^a\lambda\equiv\bar{\psi}_i\gamma^a\lambda^j(\vec{\tau})^i{}_j\label{conv1}
\end{equation}
with $(\vec{\tau})^i{}_j$ being the standard Pauli matrices and the arrow generally denotes the vector representation of $SU(2)$.   
A basic and rather useful identity is
\[
\bar\psi_i\gamma^{a_1}\ldots\gamma^{a_n}\lambda^j=-\varepsilon_{ik}\varepsilon^{jl}\bar{\lambda}_l\gamma^{a_n}\ldots\gamma^{a_1}\psi^k,
\]
from which one easily deduces the symmetry of fermion bilinears:
\[
\bar{\lambda}\gamma^{(n)}\psi=\alpha_n\bar{\psi}\gamma^{(n)}\lambda, \qquad \alpha_n= \left\{ \begin{array}{ll}
  -1&\mbox{ for }n=0,1,4,5 \\
  +1&\mbox{ for }n=2,3
\end{array}\right.,
\]
and for an additional Pauli matrix, as for example in the second formula of (\ref{conv1}), an additional minus sign has to be included.
For four Spinors $\psi,\lambda,\chi,\xi$ the Fierz identity is
\[
(\bar{\psi}_iM\lambda^j)(\bar{\chi}_kN\xi^l)=-\frac{1}{4}(\bar{\chi}_k\lambda^j)(\bar{\psi}_iMN\xi^l)-\frac{1}{4}(\bar{\chi}_k\gamma_a\lambda^j)(\bar{\psi}_iM\gamma^aN\xi^l)+\frac{1}{8}(\bar{\chi}_k\gamma_{ab}\lambda^j)(\bar{\psi}_iM\gamma^{ab}N\xi^l),
\]
where $M$ and $N$ are arbitrary (combinations of) $\gamma$-matrices.

Our spin-connection may be expressed as
\[
\omega_{mab}=\Omega_{mba}+\Omega_{abm}-\Omega_{mab}
\]
with the coefficents of anholonomy
\[
\Omega_{mn}{}^a=\partial_{[n}e_{m]}^a.
\]

The definition of supercovariance is standard: The supersymmetry transformation of a supercovariant object contains no derivatives of the transformation parameter. This is achieved by adding appropriate combinations of gravitinos. Supercovariance is indicated by a hat on the quantity in question. For example, the supercovariant graviphoton field strength is
\begin{equation}
\hat{F}_{mn}=F_{mn}+\frac{\sqrt{3}i\kappa}{2}\bar{\psi}_m\psi_n\qquad \mbox{with}\qquad F_{mn}=\partial_m A_n-\partial_n A_m.\label{fs}
\end{equation}
Another example is the  supercovariantized spin connection,
\[
\hat{\omega}_{mab}=\omega_{mab}+\frac{i\kappa^2}{2}(\bar{\psi}_m\gamma_b\psi_a-\bar{\psi}_a\gamma_m\psi_b+\bar{\psi}_b\gamma_a\psi_m).
\]
In section \ref{seclag} we use a modified spin connection
\begin{equation}
\tilde{\omega}_{mab}=\hat{\omega}_{mab}-\frac{i\kappa^2}{4}\bar{\psi}_c\gamma^{cd}{}_{mab}\psi_d.\label{spco1}
\end{equation}
Our Riemann tensor is
\[
R(\omega)^{ab}{}_{mn}=2\partial_{[m}\omega_{n]}{}^{ab}+2\omega_{[m}{}^{ac}\omega_{n]c}{}^b.
\]
Finally, for convenience, we give the supercovariant derivatives for various fields as they appear in eqns. (\ref{ftl}):
\begin{subequations}
\begin{eqnarray}
\hat{\cal D}_m\chi & = & {\cal D}_m\chi + \frac{\kappa}{2}\psi_m\varphi-3i\kappa\vec{\tau}\psi_m\vec{t}-\frac{\kappa}{2}\gamma^a\psi_m V_a\nonumber\\ & + & \frac{i\kappa}{4}\gamma^n\vec{\tau}\psi_m\vec{V}_n'-\frac{\kappa}{2}\gamma^{ab}\psi_m v_{ab}' - i\kappa^2\chi\bar{\chi}\psi_m\\
\hat{\cal D}_m V_a & = & {\cal D}_m V_a - i\kappa\bar{\psi}_m\gamma_a{}^b\hat{\cal D}_b\chi+\frac{\kappa^2}{2}\bar{\psi}_m\vec{\tau}\chi\vec{V}_a'+2i\kappa\bar{\psi}_m\gamma_a\lambda' - i\kappa^2\bar{\psi}_m\gamma^b\chi v_{ab}' - \frac{i\kappa^2}{2\sqrt{3}}\bar{\psi}_m\gamma^b\chi\hat{F}_{ab}\nonumber\\ & - & 2\kappa^2\bar{\psi}_m\vec{\tau}\gamma_a\chi\vec{t}-\frac{i\kappa^2}{2\sqrt{3}}\bar{\psi}_m\gamma_{abc}\chi\hat{F}^{bc}+i\kappa^2\bar{\psi}_m\gamma_a\gamma^b\chi V_b+i\kappa^2\bar{\psi}_m\gamma_a\chi\varphi\\
\hat{\cal D}_m v^{'ab} & = & {\cal D}_m v^{'ab} +\frac{i\kappa}{4}\bar{\psi}_m\gamma^{[a}\gamma^{b]cd}\hat{\cal R}_{cd}-\frac{3i\kappa}{2}\bar{\psi}_m\gamma^{[b}\gamma_c\hat{\cal R}^{a]c}-2i\kappa\bar{\psi}_m\gamma^{ab}\lambda'\\
\hat{\cal D}_m \vec{V}_a' & = & {\cal D}_m \vec{V}_a' +4\kappa\bar{\psi}_m\vec{\tau}\gamma_a\lambda'-\kappa\bar{\psi}_m\vec{\tau}\gamma_a{}^{bc}\hat{\cal R}_{bc}-2\kappa\bar{\psi}_m\vec{\tau}\hat{\cal D}_a\chi - i\kappa^2\bar{\chi}\psi_m\vec{V}_a' + \kappa^2\bar{\chi}\vec{\tau}\gamma_{abc}\psi_m v^{'bc}\nonumber\\ & + & \frac{\kappa^2}{2\sqrt{3}}\bar{\chi}\vec{\tau}\gamma_{abc}\psi_m\hat{F}^{bc} + 4i\kappa^2\bar{\chi}\vec{\tau}\tilde{\tau}\gamma_a\psi_m\tilde{t}-\frac{2\kappa^2}{\sqrt{3}}\bar{\chi}\vec{\tau}\gamma^b\psi_m\hat{F}_{ab}+\kappa^2\bar{\psi}_m\vec{\tau}\chi\times\vec{V}_a'\\
\hat{\cal D}_m \vec{t} & = & \partial_m \vec{t} - \kappa\bar{\psi}_m\vec{\tau}\lambda'+\frac{\kappa}{4}\bar{\psi}_m\vec{\tau}\gamma^{ab}\hat{\cal R}_{ab} - 2\kappa^2\vec{t}\times(\bar{\psi}_m\vec{\tau}\chi).
\end{eqnarray}\label{covder}
\end{subequations}
\end{appendix}
We recall that the covariant derivatives appearing here are only covariant w.r.t. local Lorentz transformations.

\end{document}